\documentclass[11pt]{article}
\usepackage[margin=1in]{geometry}
\usepackage[T1]{fontenc}
\usepackage{lmodern}
\usepackage{microtype}
\usepackage{graphicx}
\usepackage{amsmath,amsfonts,amssymb}
\usepackage{bm}
\usepackage{float}
\usepackage[hidelinks]{hyperref}

\begin{document}

\title{Coulomb- and exchange-driven redistribution of spectral correlations in $\gamma$-Ce}

\author{Zanya Gonz\'alez-T\'ellez$^1$ and M. A. Oca\~na-Bribiesca$^{1,*}$\\[0.6em]
\small $^1$Departamento de Actuar\'ia, F\'isica y Matem\'aticas, Universidad de las Am\'ericas Puebla,\\ Ex-hacienda Sta. Catarina M\'artir s/n, San Andr\'es Cholula, Pue. 72810, M\'exico\\
\small $^*$Author to whom any correspondence should be addressed.\\
\small \texttt{miguel.ocana@udlap.mx}}

\date{}
\maketitle

% --------------------------------------------- ABSTRACT ----------------------------------------------------------------

\begin{abstract}

Electronic correlations reorganize not only the energies of individual electronic states but also the collective structure of the spectrum. Here, first-principles density functional theory calculations combined with random-matrix spectral statistics are used to examine this reorganization in $\gamma$-Ce within a static spin-polarized DFT+$U$+SOC description. We distinguish the effects of the Coulomb interaction $U$ from those of the exchange interaction $J$ by analyzing nearest-neighbor spacings, reduced spacing ratios, band velocities, and band curvatures across the Ce-$4f$ manifold. Increasing $U$ does not produce a monotonic transition between Poisson- and Wigner--Dyson-like statistics; instead, the region of strongest level repulsion migrates across the $4f$ manifold as the interaction increases. Exchange produces a qualitatively different response. At fixed $U$, strongly repulsive and strongly clustered spectral regions coexist within the same $4f$ manifold even while the total spin moment and $4f$ occupation remain nearly unchanged. The clustering persists after resolving the spectrum into its dominant $j=5/2$ and $j=7/2$ components and differs strongly between calculations having the same numerical value of $U-J$. Within the present DFT+$U$+SOC description, the spectral response therefore depends on the separate Coulomb and exchange interactions and cannot be represented completely by a single effective $U-J$ parameter. These results show that energy- and character-resolved spectral statistics expose correlation-driven electronic reorganization that is hidden by conventional integrated observables.
\end{abstract}


% ------------------------------------ INTRODUCCION -----------------------------------------------------------------

\section{Introduction}\label{Introduction}

Strongly correlated electron systems present a major challenge in condensed matter physics due to the significant role played by electron-electron interactions. These interactions give rise to a wide variety of physical phenomena including metal-insulator transitions, heavy-fermion behavior, unconventional superconductivity, and magnetic ordering \cite{imada1998metal,georges1996dynamical}. The description of these systems requires computational methods capable of reproducing the electronic structure while incorporating correlation effects that extend beyond conventional independent-particle approximations.

Density Functional Theory (DFT) has become one of the most successful first-principles approaches for calculating the electronic structure of solids \cite{jones2015density}. Within the Local Density Approximation (LDA) and the Generalized Gradient Approximation (GGA), DFT has successfully described the electronic properties of semiconductors, metals, and a broad range of crystalline materials. However, conventional DFT often fails to describe materials containing localized $d$ and $f$ electrons, where the Coulomb interaction is sufficiently strong to modify the electronic spectrum. Various extensions have therefore been developed, including the DFT+$U$ method \cite{anisimov1991hubbard,anisimov1997lda}, hybrid functionals \cite{heyd2003hybrid}, and Dynamical Mean-Field Theory (DFT+DMFT) \cite{kotliar2006electronic,haule2010dynamical,maier2005quantum}. These approaches improve the description of electron-electron interactions using different approximations while maintaining the advantages of first-principles calculations.

Electronic structure calculations are commonly analyzed through quantities such as band structures, densities of states, magnetic moments, or charge densities. These observables describe specific features of the electronic structure, but they do not directly characterize the collective organization of the electronic spectrum. Since electron correlation modifies the relative position of electronic bands, changes the degree of localization of electronic states, and enhances level repulsion, it is natural to expect that these effects should also be reflected in the statistical properties of the spectrum. Statistical descriptors therefore provide a complementary perspective on the evolution of correlated electronic systems.

Random Matrix Theory (RMT) provides a natural framework for studying spectral statistics in complex quantum systems \cite{mehta2004random,haake2010quantum}. Originally developed to describe the spectra of atomic nuclei, RMT has become one of the standard tools for investigating quantum chaos. According to this theory, integrable systems are characterized by Poisson level-spacing statistics, whereas chaotic systems exhibit Wigner--Dyson level repulsion described by the Gaussian Orthogonal Ensemble (GOE), Gaussian Unitary Ensemble (GUE), or Gaussian Symplectic Ensemble (GSE), depending on the symmetries of the system \cite{mehta2004random,haake2010quantum}. In the present analysis, Poisson, GOE, and GUE statistics are used as reference distributions for the calculated spectra rather than as exact symmetry assignments of the effective Hamiltonian.

The application of RMT to electronic band structures was demonstrated by Mucciolo et al.~\cite{mucciolo1996manifestation}, who showed that first-principles electronic bands may exhibit universal spectral statistics in selected regions of the Brillouin zone. Related transitions between Poisson, GOE, and GUE statistics have also been observed in other quantum systems, including magnetoexcitons subjected to external magnetic fields \cite{schweiner2017goe} and quantum spin chains with random interactions \cite{kundu2023spectral}.
%The application of RMT to electronic band structures was demonstrated by Mucciolo \textit{et al.} \cite{mucciolo1996manifestation}, who showed that first-principles electronic bands may exhibit universal spectral statistics in selected regions of the Brillouin zone. Since then, similar transitions between Poisson, GOE, and GUE statistics have been observed in a variety of quantum systems, including magnetoexcitons subjected to external magnetic fields \cite{schweiner2017goe} and quantum spin chains with random interactions \cite{kundu2023spectral}. These studies indicate that spectral statistics provide valuable information about the complexity of the underlying quantum system. Beyond quantum chaos, Random Matrix Theory has become a useful framework for describing universal spectral fluctuations in complex systems. In condensed matter physics, RMT has been successfully applied to disordered conductors, mesoscopic systems, and interacting many-body systems, where microscopic details differ but universal spectral correlations emerge. This universality makes spectral statistics particularly attractive for the analysis of electronic spectra obtained from first-principles calculations, since they provide information that complements conventional electronic-structure observables.

Among strongly correlated materials, Cerium provides a particularly suitable model system because its electronic behavior is governed by a single $4f$ electron. The competition between localization of the $4f$ electron and its hybridization with the conduction bands makes its electronic structure highly sensitive to local electronic correlations. Consequently, the $\gamma$ phase of Cerium is an excellent platform for investigating how increasing electronic correlation modifies the statistical organization of the electronic spectrum \cite{allen1982,shick2001cerium,kotliar2006electronic}.

From this perspective, strongly correlated materials constitute an interesting scenario for the application of Random Matrix Theory. Increasing the Hubbard interaction modifies the electronic spectrum through changes in band localization, hybridization, and level repulsion. Previous correlated band-theory calculations have shown that relatively small changes in the treatment of local electronic correlations can produce distinct self-consistent electronic solutions in $\gamma$-Ce \cite{shick2001cerium}. This sensitivity raises a more specific question: whether the Coulomb and exchange components of the local interaction reorganize the electronic spectrum in the same manner. Conventional observables such as the total magnetic moment, orbital occupation, or density of states may remain similar even when the relative organization of individual electronic levels changes substantially. Spectral statistics provide a way of resolving this internal organization by quantifying level repulsion, clustering, and the dynamics of neighboring bands. The question is therefore whether $U$ and $J$ generate distinct forms of spectral reorganization within the correlated $4f$ manifold.

In this work, we investigate how the Coulomb and exchange components of the local interaction separately reorganize the effective one-electron spectrum of $\gamma$-Ce within a static spin-polarized DFT+$U$+SOC description. Two interaction paths are considered: a $J=0$ series with variable $U$, and a fixed-$U=6.12$ eV series with variable $J$. Because the position and width of the high-energy Ce-$4f$ manifold evolve with the interaction parameters, the relevant spectral region is identified independently for each calculation from the projected density of states. We analyze nearest-neighbor spacings, reduced spacing ratios, band velocities, and band curvatures for the complete $4f$ manifold, for energy-resolved subwindows, and for states with dominant $j=5/2$ and $j=7/2$ character. The results reveal qualitatively different responses to $U$ and $J$. Increasing $U$ redistributes the region of strongest level repulsion across the $4f$ manifold rather than producing a uniform statistical crossover, whereas increasing $J$ generates a coexistence of strongly repulsive and strongly clustered spectral regions. The exchange-driven clustering persists within the individual relativistic subsequences and differs substantially between spectra having the same numerical value of $U-J$. These results demonstrate that the internal organization of the correlated $4f$ spectrum contains information about the separate roles of Coulomb and exchange interactions that is not captured by conventional integrated observables or by a single effective interaction parameter.

% ---------------------------------------- MODEL & COMPUTATIONAL DETAILS ---------------------------------------------

\section{Numerical Methodology}

\subsection{Electronic Structure Calculations}

The calculations were performed for the {\it fcc} structure of $\gamma$-Ce. Before the electronic-structure calculations, the equilibrium lattice parameter was determined from total-energy calculations as a function of volume. The optimized lattice constant was $a=5.2803$~\AA, approximately 2.31\% larger than the reported experimental value of $5.1610$~\AA~\cite{beaudry1974lattice}, and was used throughout this work.% The optimized lattice constant was $a_0=5.2803$~\AA. For the DFT+$U$+SOC calculations analyzed in this work, the lattice parameter was subsequently expanded to $a=5.5442$~\AA, corresponding to an increase of approximately 5\%. The expanded structure was used to reduce the hybridization of the Ce-$4f$ states with the more itinerant bands and to enhance the localized character of the $4f$ manifold considered in the statistical analysis.

Electronic-structure calculations were performed within Density Functional Theory (DFT) using the full-potential linearized augmented plane-wave (FP-LAPW) method implemented in the WIEN2k package \cite{blaha2001wien2k}. Exchange and correlation were treated using the Wu-Cohen generalized gradient approximation (WC-GGA) \cite{wucohen}. Spin polarization, spin-orbit coupling, and local electronic correlations within the Ce-$4f$ shell were included. Local correlations were treated using the rotationally invariant DFT+$U$ formalism \cite{anisimov1997lda,shick1999implementation}. The self-interaction correction (SIC) was used for the double-counting term. Spin-orbit coupling was subsequently included using the second-variational method, while the complex orbital-dependent potential was retained in the spin-orbit-coupled calculation. %The plane-wave cutoff was defined by $RK_{\max}=9.0$. The Brillouin-zone mesh was generated using a target of 2000 $k$-points, and an electronic temperature of $0.0272$ eV ($0.002$ Ry) was used in the self-consistent calculations. (after the Wu-Cohen cite)

The calculations were carried out in three consecutive stages. First, a spin-polarized DFT calculation was converged. The resulting electronic density was then used to initialize the DFT+$U$ calculation, and spin-orbit coupling was included after convergence of the DFT+$U$ stage. The post-spin-orbit eigenvalues obtained from the final DFT+$U$+SOC calculation were used in the statistical analysis.

Two families of electronic spectra were generated. In the first series, the exchange parameter was fixed at $J=0.0$ eV, while $U$ was varied from $0.0$ to $9.52$ eV ($0.0$--$0.70$ Ry) in steps of approximately $0.68$ eV ($0.05$ Ry). In the second series, $U$ was fixed at $6.12$ eV ($0.45$ Ry), while $J$ was varied from $0.0$ to $0.95$ eV ($0.0$--$0.07$ Ry) in steps of approximately $0.14$ eV ($0.01$ Ry). For the first series, $J=0.0$ and the effective interaction $U-J$ therefore reduces numerically to $U$, corresponding to the $U_{\mathrm{eff}}$ description commonly used in the simplified rotationally invariant approach of Dudarev \textit{et al.} \cite{Dudarev1998}. This series therefore provides a direct measure of the evolution of the spectrum with the effective Coulomb interaction. The second series was constructed at fixed $U$ while varying $J$, allowing the effect of the exchange interaction to be examined independently. This separation is particularly relevant for $\gamma$-Ce, where correlated band calculations have shown that the balance between Coulomb and exchange interactions plays an important role in determining the $4f$ electronic configuration and the resulting magnetic solutions \cite{shick2001cerium}. The sensitivity of the electronic structure of fcc Cerium to the treatment of the $4f$ states and local electronic correlations has also been demonstrated in one-electron calculations by Tran \textit{et al.} \cite{TranBlaha2014}. 

Because the DFT+$U$ functional may possess multiple self-consistent solutions for the Ce-$4f$ shell, the sensitivity of the calculations to the initialization procedure was examined using converged neighboring DFT+$U$ solutions as alternative starting states before the inclusion of spin-orbit coupling. When distinct converged solutions were obtained at the same interaction parameters, their post-SOC total energies were compared and the lower-energy solution was retained for the reported production values. A separate robustness test was performed with respect to reciprocal-space sampling by repeating the short-range energy- and relativistic-character-resolved statistical analysis along the $L\rightarrow K$ path, using the same DOS-based window definition, 1001-point path sampling, and interior interval $k=50$--950. The $W\rightarrow L$ trajectory was retained as the primary path for the analysis presented below, whereas the $L\rightarrow K$ calculations were used to assess the robustness of the qualitative spectral trends with respect to reciprocal-space direction.

\subsection{Spectral Statistical Analysis}

The statistical analysis was performed using the post-spin-orbit eigenvalues obtained from the final self-consistent calculations. The analysis follows the Random Matrix Theory framework commonly used to characterize short-range correlations in quantum spectra \cite{mehta2004random,haake2010quantum}. A similar statistical treatment of eigenvalues and band velocities has previously been applied to first-principles electronic band structures \cite{mucciolo1996manifestation}.

Because the position and width of the Ce-$4f$ states change with the interaction parameters, the same fixed energy interval was not used for every case. Instead, the relevant spectral region was identified independently from the Ce-$4f$ projected density of states. For each calculation, the high-energy $4f$ spectral window in the density of states was identified using a threshold equal to $2\%$ of the maximum $4f$ spectral weight in the search interval. Components separated by less than $0.10$ eV were merged, while components containing less than $5\%$ of the selected $4f$ spectral weight were discarded. The lower and upper limits of the selected component were determined from the $1$st and $99$th percentiles of its cumulative spectral weight, and the resulting interval was extended by $0.10$ eV on each side. This procedure allowed the selected energy window to follow the displacement and broadening of the relevant Ce-$4f$ manifold as $U$ and $J$ were varied.

In addition to the complete high-energy $4f$ manifold interval, three overlapping subwindows were defined using the cumulative integrated $4f$ density of states. The lower window contained the first 0--40\% of the integrated $4f$ spectral weight, the central window contained 30--70\%, and the upper window contained 60--100\%. The overlap between neighboring windows was introduced deliberately in order to reduce the dependence of the results on a single sharp energy boundary and to allow changes in the statistical behavior to be followed through different regions of the $4f$ manifold.

The electronic bands were evaluated along the direct $W\rightarrow L$ path using 1001 uniformly spaced reciprocal-space points. The statistical analysis was restricted to points $k=50$--$950$, corresponding to 901 points after removing the ends of the path to avoid high-symmetry boundary degeneracies. At every $k$-point, the eigenvalues within the selected energy window were ordered in increasing energy.

Consecutive level spacings were calculated as
\begin{equation}
\Delta_n(k)=E_{n+1}(k)-E_n(k).
\end{equation}
The spacings were normalized by the mean spacing of the selected levels at the same $k$-point,
\begin{equation}
s_n(k)=
\frac{\Delta_n(k)}
{\langle\Delta(k)\rangle},
\end{equation}
where $\langle\Delta(k)\rangle$ is the local mean level spacing. This local normalization removes variations in the energy scale caused by changes in the width and density of the selected spectrum, without requiring a global unfolding \cite{mehta2004random,haake2010quantum,mucciolo1996manifestation}.% It does not involve a polynomial, spline, or global unfolding of the electronic spectrum.

The normalized spacings were used to construct the nearest-neighbor spacing distribution, $P(s)$, and the numerical distributions were compared with the Poisson, Gaussian Orthogonal Ensemble (GOE), and Gaussian Unitary Ensemble (GUE) predictions \cite{mehta2004random,haake2010quantum}. The corresponding reference distributions are
\begin{equation}
P_{\mathrm{P}}(s)=\exp(-s),
\end{equation}
\begin{equation}
P_{\mathrm{GOE}}(s)=
\frac{\pi}{2}s
\exp\left(-\frac{\pi}{4}s^2\right),
\end{equation}
and
\begin{equation}
P_{\mathrm{GUE}}(s)=
\frac{32}{\pi^2}s^2
\exp\left(-\frac{4}{\pi}s^2\right).
\end{equation}
The Poisson distribution describes spectra without level repulsion, whereas the GOE and GUE distributions exhibit linear and quadratic level repulsion, respectively \cite{mehta2004random,haake2010quantum}.

The spacing distributions were also fitted using the Brody distribution \cite{brody1981random},
\begin{equation}
P_{\mathrm{B}}(s)=
(\beta_{\mathrm{B}}+1)b
s^{\beta_{\mathrm{B}}}
\exp\left[-b s^{\beta_{\mathrm{B}}+1}\right],
\end{equation}
where
\begin{equation}
b=
\left[
\Gamma\left(
\frac{\beta_{\mathrm{B}}+2}
{\beta_{\mathrm{B}}+1}
\right)
\right]^{\beta_{\mathrm{B}}+1}.
\end{equation}
The Brody parameter varies from $\beta_{\mathrm{B}}=0$ for Poisson statistics to $\beta_{\mathrm{B}}=1$ for GOE-like level repulsion \cite{brody1981random}. Since the conventional Brody distribution does not distinguish between GOE and GUE statistics, the value of $\beta_{\mathrm{B}}$ was interpreted together with the spacing-ratio statistics and the shape of the numerical distributions.

As a complementary measure of short-range spectral correlations, the reduced consecutive-spacing ratio was calculated as \cite{atas2013distribution}
\begin{equation}
r_n(k)=
\frac{
\min\left[\Delta_n(k),\Delta_{n+1}(k)\right]
}{
\max\left[\Delta_n(k),\Delta_{n+1}(k)\right]
}.
\end{equation}
This ratio satisfies $0\leq r\leq1$. Because the local energy scale cancels in the ratio, no additional normalization or unfolding is required. The expected average values for Poisson, GOE, and GUE statistics are $\langle r\rangle_{\mathrm{P}}\simeq0.3863$, $\langle r\rangle_{\mathrm{GOE}}\simeq0.5359$, and $\langle r\rangle_{\mathrm{GUE}}\simeq0.6027$, respectively.
%\begin{equation}
%\langle r\rangle_{\mathrm{P}}\simeq0.3863,
%%\qquad
%\langle r\rangle_{\mathrm{GOE}}\simeq0.5359,
%%\qquad
%\langle r\rangle_{\mathrm{GUE}}\simeq0.6027.
%\end{equation}

The nearest-neighbor spacing and reduced-ratio distributions were calculated for the complete  high-energy $4f$ spectral interval and for the lower, central, and upper energy subwindows. This energy-resolved analysis was used to determine whether the statistical behavior of the complete manifold was uniform or resulted from the coexistence of different spectral regimes. To determine whether the spectral behavior also depends on relativistic character, the upper $4f$ manifold was analyzed separately in terms of its $j=5/2$ and $j=7/2$ components. The corresponding energy intervals were identified independently from the total $j$-projected Ce-$4f$ partial densities of states. The post-SOC contributions from the two WIEN2k spin channels were summed for this purpose, so that the resulting curves represent the total $j$-projected $4f$ partial density of states rather than separate spin-up and spin-down spectra. For each state, the total projected $4f$ weight was defined as $w_f=w_{5/2}+w_{7/2}$. States with $w_f\geq0.05$ were assigned to the $j=5/2$ sector when $w_{5/2}/w_f\geq0.60$, and to the $j=7/2$ sector when $w_{7/2}/w_f\geq0.60$. The level spacings, local normalization, reduced spacing ratios, and Brody fits were then evaluated independently for the two character-selected subsequences using the same reciprocal-space interval $k=50$--$950$.

To connect the spectral statistics with the evolution of the electronic bands, the first and second derivatives of the energy-ordered eigenvalues were calculated numerically along the $W\rightarrow L$ path. The use of crystal momentum as the parameter controlling the evolution of the bands follows the level-dynamics approach previously applied to first-principles band structures \cite{mucciolo1996manifestation}. Band-velocity statistics characterize the distribution of local band slopes and provide information about the dispersion of the states along the reciprocal-space path. Band-curvature statistics describe the local bending and reorganization of neighboring bands \cite{zakrzewski1993parametric}. Using the physical reciprocal-space coordinate $q$, the band slope was defined as
\begin{equation}
v_n(q)=\frac{dE_n}{dq},
\end{equation}
and the band curvature as
\begin{equation}
C_n(q)=\frac{d^2E_n}{dq^2}.
\end{equation}
The first derivative is proportional to the electronic group velocity, while the second derivative describes the local bending of the electronic band. The corresponding physical quantities have units of eV\,\AA\ and eV\,\AA$^2$, respectively.

For comparison among different values of $U$ and $J$, the derivatives were also scaled using the normalized path coordinate $x$ and the local mean level spacing,
\begin{equation}
\widetilde v_n(k)=
\frac{dE_n/dx}
{\langle\Delta(k)\rangle},
\end{equation}
and
\begin{equation}
\widetilde C_n(k)=
\frac{d^2E_n/dx^2}
{\langle\Delta(k)\rangle}.
\end{equation}
This scaling expresses the band motion and band bending relative to the local spectral separation. As in the level-spacing analysis, this procedure represents a dimensionless spectral scaling and not a global unfolding.

The signed scaled-velocity distributions were compared with Gaussian distributions characterized by their fitted mean and standard deviation. Skewness and excess kurtosis were used to describe asymmetry and deviations from a Gaussian shape. The Gaussian fit was used only as a descriptive measure of the velocity distribution and did not assume that the individual band velocities were statistically independent.

Since positive and negative curvatures cancel in an ordinary average, the curvature analysis was based primarily on the magnitude $|\widetilde C|$. The median curvature magnitude was used to represent the typical curvature scale. The fraction of samples with a curvature magnitude greater than five times the local median was used to identify the contribution of unusually sharp curvature events.

For visualization of the curvature distributions, the magnitude was additionally normalized by the median value within the selected spectral window,
\begin{equation}
\widetilde c =
\frac{|\widetilde C|}
{\mathrm{median}(|\widetilde C|)}
\end{equation}
With this definition, $\widetilde c=1$ corresponds to the median scaled-curvature magnitude of the selected spectrum. The distributions shown in Fig.~\ref{fig:band_dynamics_J005} are expressed as $P(\widetilde c)$.

Adjacent-band gaps were also followed along the reciprocal-space path. A small-gap minimum was identified when an adjacent gap formed a local minimum with respect to $k$ and its value was smaller than $0.25$ times the local mean level spacing. These minima were used as an additional measure of the occurrence of near-degeneracies and possible crossing-related behavior.

The derivatives were calculated using energy-ordered eigenvalue indices rather than wavefunction-overlap tracking. Consequently, the largest curvature events may include contributions from avoided crossings, exact crossings, or exchanges in the energy ordering of neighboring bands. For this reason, the median curvature was used as the principal measure of the curvature scale, while the extreme tail was interpreted as an indicator of crossing and band-reordering activity.

% ------------------------------------- RESULTS AND DISCUSSION --------------------------------------------

\section{Results and Discussion}\label{Sec:Results}

\subsection{Evolution of the Electronic Structure with Local Electronic Correlations}

The electronic-structure quantities and DOS-selected spectral intervals obtained along the two interaction paths are summarized in Table~\ref{tab:dft_observables}. The table includes the total spin magnetic moment, the Ce-$4f$ occupation, the energy interval associated with the high-energy $4f$ manifold selected for the statistical analysis, and the full-window reduced spacing ratio and Brody parameter. %The point $U=6.12$ eV and $J=0.0$ is common to both interaction series and is reported only once.

\begin{table}[t]
\centering
\caption{Electronic-structure quantities, DOS-selected spectral intervals, and full-window spectral statistics obtained along the two interaction paths. $M_{\mathrm{tot}}$ is the total spin magnetic moment per unit cell, $N_{4f}$ is the total Ce-$4f$ occupation, and the high-energy $4f$ window gives the DOS-selected energy interval relative to the Fermi energy. $\langle r\rangle_{\mathrm{full}}$ and $\beta_{\mathrm{B}}$ are the mean reduced spacing ratio and Brody parameter calculated for the high-energy $4f$ spectral window, respectively. The point $U=6.12$ eV, $J=0.0$ eV is common to both series.}
\label{tab:dft_observables}
\small
\setlength{\tabcolsep}{3.5pt}
\begin{tabular}{ccccccc}
\hline
$U$ (eV) & $J$ (eV) & $M_{\mathrm{tot}}$ ($\mu_B$) & $N_{4f}$ & High-energy $4f$ window (eV) & $\langle r\rangle_{\mathrm{full}}$ & $\beta_{\mathrm{B}}$ \\
\hline
\multicolumn{7}{c}{$J=0.0$, variable $U$} \\
\hline
0.00 & 0.00 & 1.48034 & 1.02504 & $[-0.45,\;1.74]$ & 0.5042 & 0.4275 \\
0.68 & 0.00 & 1.35842 & 1.01886 & $[-0.70,\;2.19]$ & 0.4738 & 0.0877 \\
1.36 & 0.00 & 1.20100 & 0.99735 & $[-0.05,2.45]$ & 0.4992 & 0.3019 \\
2.04 & 0.00 & 1.12453 & 0.98178 & $[0.32,\;2.76]$ & 0.4188 & 0.1999 \\
2.72 & 0.00 & 1.11236 & 0.97417 & $[0.49,\;3.03]$ & 0.4684 & 0.2326 \\
3.40 & 0.00 & 1.11736 & 0.96939 & $[-0.37,\;2.27]$ & 0.4608 & 0.3246 \\
4.08 & 0.00 & 1.11923 & 0.96673 & $[1.02,\;3.61]$ & 0.5577 & 0.5274 \\
4.76 & 0.00 & 1.12927 & 0.96502 & $[1.62,\;3.95]$ & 0.5265 & 0.5510 \\
5.44 & 0.00 & 1.12800 & 0.96419 & $[1.83,\;4.31]$ & 0.4964 & 0.5642 \\
6.12 & 0.00 & 1.13312 & 0.96379 & $[2.08,\;4.57]$ & 0.4996 & 0.5615 \\
6.80 & 0.00 & 1.14271 & 0.96378 & $[2.28,\;4.88]$ & 0.4861 & 0.5395 \\
7.48 & 0.00 & 1.13440 & 0.96393 & $[2.69,\;5.16]$ & 0.5029 & 0.7579 \\
8.16 & 0.00 & 1.14178 & 0.96422 & $[3.15,\;5.47]$ & 0.4809 & 0.5425 \\
8.84 & 0.00 & 1.13312 & 0.96455 & $[3.35,\;5.79]$ & 0.4617 & 0.2995 \\
9.52 & 0.00 & 1.11758 & 0.96492 & $[3.65,\;6.13]$ & 0.5105 & 0.3634 \\
\hline
\multicolumn{7}{c}{$U=6.12$ eV, variable $J$} \\
\hline
6.12 & 0.14 & 1.13199 & 0.96409 & $[2.05,\;4.60]$ & 0.4838 & 0.4913 \\
6.12 & 0.27 & 1.14313 & 0.96446 & $[1.98,\;4.63]$ & 0.4932 & 0.4641 \\
6.12 & 0.41 & 1.13044 & 0.96468 & $[1.89,\;4.68]$ & 0.5083 & 0.4241 \\
6.12 & 0.54 & 1.13186 & 0.96514 & $[1.84,\;4.70]$ & 0.4576 & 0.3316 \\
6.12 & 0.68 & 1.13385 & 0.96572 & $[1.76,\;4.74]$ & 0.4122 & 0.1927 \\
6.12 & 0.82 & 1.13211 & 0.96636 & $[1.67,\;4.77]$ & 0.3919 & 0.1689 \\
6.12 & 0.95 & 1.13428 & 0.96705 & $[1.59,\;4.79]$ & 0.4302 & 0.2023 \\
\hline
\end{tabular}
\end{table}

The inclusion of local electronic correlations produces clear changes in the position and internal organization of the Ce-$4f$ states. However, the effects of the Coulomb interaction $U$ and the exchange parameter $J$ are different. Increasing $U$ produces a progressive displacement of the upper $4f$ manifold toward higher energies, whereas increasing $J$ at fixed $U$ produces a redistribution of the spectral weight over a broader energy range. These changes can be observed in the projected density of states and in the corresponding electronic bands.

\subsubsection{Evolution with Hubbard $U$}

The most evident effect of increasing $U$ is the displacement of the high-energy Ce-$4f$ manifold toward higher energies. The DOS-selected intervals for the complete $U$ series are reported in Table~\ref{tab:dft_observables}. At $U=0.0$, part of the high-energy $4f$ spectral weight remains close to the Fermi level. As the interaction increases, this component progressively moves toward higher energies, with the displacement becoming particularly clear for $U\gtrsim4.08$ eV.

The evolution is not equivalent to a rigid shift of an unchanged set of bands. The distribution of the $4f$ spectral weight, the relative separation of the bands, and their dispersion also change with $U$. At the smaller interaction values, the structure of the projected density of states changes sufficiently that the boundaries of the selected component do not evolve monotonically. Above approximately $U=4.08$ eV, however, the displacement toward higher energies becomes clearly established.

The conventional electronic-structure observables also reflect a reorganization at the smaller values of $U$. The total spin magnetic moment is $1.48034~\mu_B$ at $U=0.0$ and changes considerably up to approximately $U=2.04$ eV. For larger values of $U$, it becomes substantially more stable, remaining close to $1.1$--$1.14~\mu_B$ throughout most of the series. A similar stabilization is observed in the $4f$ occupation. Starting from $N_{4f}=1.02504$ at $U=0.0$, the occupation approaches approximately $0.96$--$0.97$ for the intermediate and large values of $U$.

For the spectral analysis, the important consequence is that the region containing the high-energy $4f$ states moves considerably in energy as $U$ changes. A single fixed energy interval would therefore sample different parts of the $4f$ manifold at different interaction strengths. This provides the motivation for the DOS-dependent selection procedure already described.

\begin{figure}[!htbp]
\centering
\includegraphics[width=\linewidth]{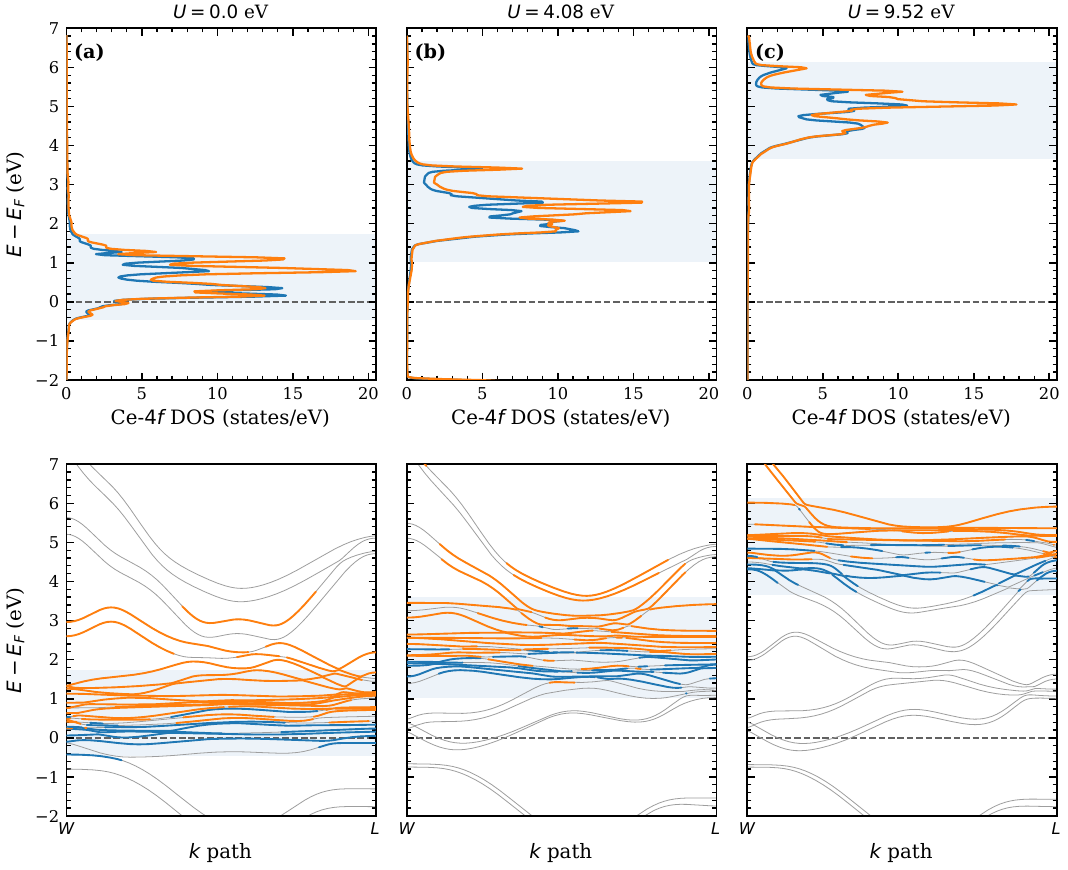}
\caption{Evolution of the electronic structure along the $U$ cut ($J = 0.0$ eV). The upper panels display the total $j$-projected Ce-$4f$ partial densities of states (pDOS) for (a) $U = 0.0$, (b) $4.08$, and (c) $9.52$ eV, with blue and orange lines representing the $j = 5/2$ and $j = 7/2$ components, respectively. The lower panels show the corresponding electronic band structures along the $W \rightarrow L$ path, where colored segments highlight states matching the dominant relativistic character. The shaded region marks the DOS-selected upper Ce-$4f$ manifold used in the spectral statistical analysis.}
\label{fig:electronic_structure_U}
\end{figure}

Figure~\ref{fig:electronic_structure_U} shows representative $j$-projected Ce-$4f$ partial densities of states and band structures at $U=0.0$, $4.08$, and $9.52$ eV. These values illustrate the initial spectrum, the intermediate region in which strong changes in the level statistics are found, and the large-$U$ regime. The selected high-energy $4f$ interval is indicated in each case.

\subsubsection{Evolution with Exchange Parameter $J$}

The influence of the exchange interaction was examined at fixed $U=6.12$ eV. The corresponding DOS-selected intervals are reported in Table~\ref{tab:dft_observables}. In contrast with the predominantly upward displacement produced by $U$, increasing $J$ redistributes the high-energy $4f$ states over a broader energy range. The lower boundary moves toward the Fermi energy, while the upper boundary changes more slowly, producing an overall broadening of the manifold.

The broadening of the $4f$ manifold is accompanied by modifications in the relative organization and dispersion of the electronic bands within the selected region. Therefore, the effect of $J$ involves not only a change in the width of the manifold, but also a reorganization of its internal electronic structure.

In contrast with these visible changes in the spectrum, the conventional integrated observables remain remarkably stable along the $J$ cut. The total spin magnetic moment remains close to $1.13~\mu_B$ throughout the series, with only small variations. The total $4f$ occupation changes from $N_{4f}=0.96379$ at $J=0.0$ to $0.96705$ at $J=0.95$ eV.

This result is important for the statistical analysis that follows. Increasing $J$ produces only small changes in the total $4f$ occupation and spin magnetic moment, while the statistical organization of the individual electronic levels changes considerably. In particular, the strongest differences in the spectral statistics occur around $J=0.68$--$0.82$ eV, where $M_{\mathrm{tot}}$ and $N_{4f}$ remain almost unchanged. The statistical properties of the spectrum therefore provide information that is not apparent from these conventional electronic-structure observables alone.

\begin{figure}[!htbp]
\centering
\includegraphics[width=\textwidth]{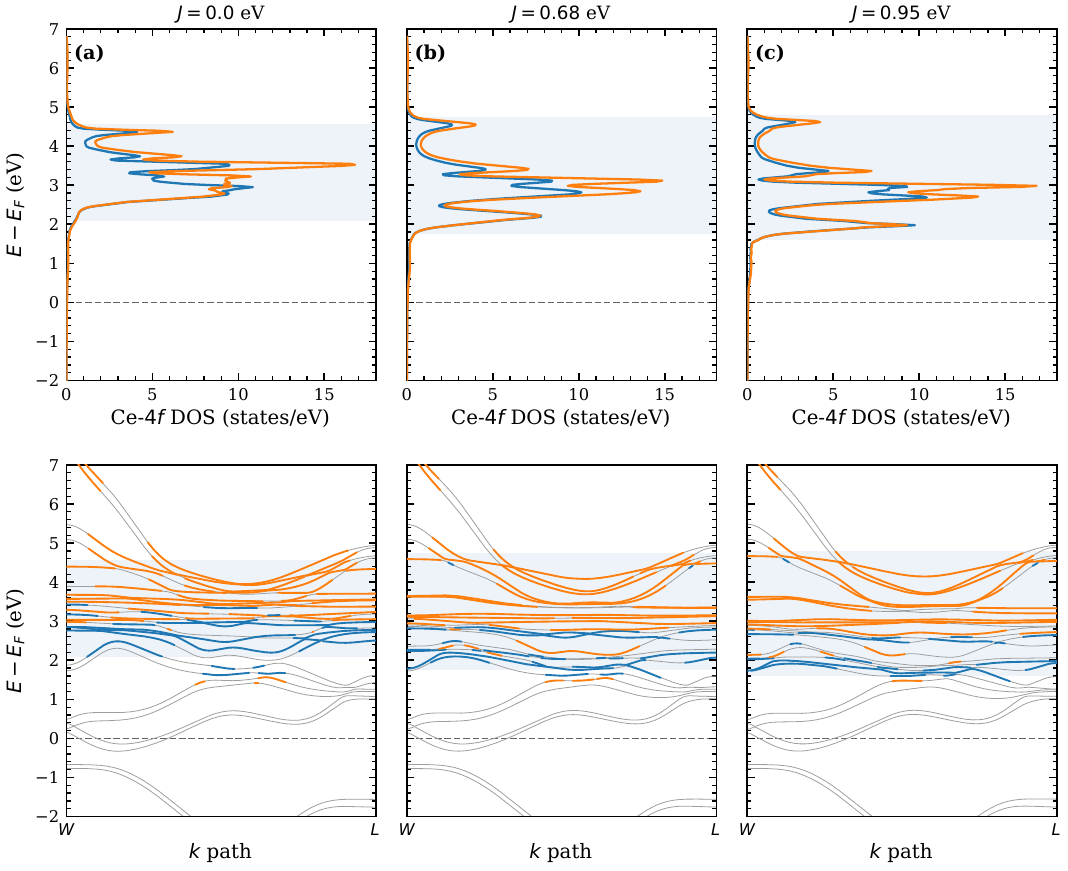}
\caption{Evolution of the electronic structure along the $J$ cut at fixed $U = 6.12$ eV. The upper panels show the total $j$-projected Ce-$4f$ partial densities of states (pDOS) for (a) $J = 0.0$, (b) $0.68$, and (c) $0.95$ eV, with blue and orange lines denoting the $j = 5/2$ and $j = 7/2$ components, respectively. The lower panels show the corresponding electronic band structures along the $W \rightarrow L$ path, with colored segments identifying states matching the dominant relativistic character. The shaded region marks the DOS-selected high-energy Ce-$4f$ manifold used in the spectral statistical analysis.}

\label{fig:electronic_structure_J}
\end{figure}

Figure~\ref{fig:electronic_structure_J} illustrates this evolution through the $j$-projected Ce-$4f$ partial densities of states and corresponding band structures for $J=0.0$, $0.68$, and $0.95$ eV. The value $J=0.68$ eV is particularly important because, as shown in the following sections, the central and upper regions of the same $4f$ manifold display strongly contrasting spectral statistics at this interaction strength.

\subsection{Statistical Analysis of the Electronic Spectrum}

The full-window spectral statistics are summarized in Table~\ref{tab:dft_observables} and already indicate that $U$ and $J$ do not simply tune the spectrum along a common Poisson--Wigner--Dyson crossover, but produce distinct forms of spectral reorganization within the Ce-$4f$ manifold. Neither interaction path shows a simple monotonic evolution between the Poisson, GOE, and GUE reference regimes. Along the $U$ cut, both $\langle r\rangle$ and the Brody parameter vary nonmonotonically as the interaction strength increases. Along the $J$ cut, the complete-window spectrum shows a clearer reduction of level repulsion, with $\langle r\rangle$ decreasing from approximately $0.50$ at $J=0.0$ to $0.392$ at $J=0.82$ eV, close to the Poisson reference value. The corresponding Brody parameter decreases from approximately $0.56$ to $0.17$.

\begin{figure}[!htbp]
\centering
\includegraphics[width=\linewidth]{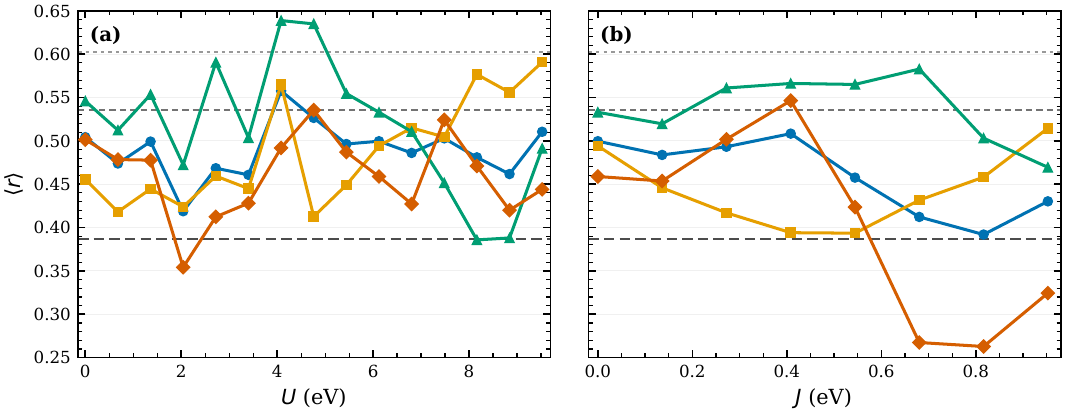}
\caption{Energy-resolved evolution of the mean reduced spacing ratio $\langle r \rangle$ as a function of (a) Hubbard $U$ (at $J = 0.0$ eV) and (b) exchange parameter $J$ (at fixed $U = 6.12$ eV) for the different DOS-selected Ce-$4f$ subwindows. Horizontal dashed lines represent the analytical Poisson, GOE, and GUE random matrix reference values.}
\label{fig:r_evolution}
\end{figure}

The complete-window quantities, however, average over the entire DOS-selected $4f$ manifold. To determine whether this behavior is uniform throughout the selected spectrum, the same analysis was performed for the lower, central, and upper DOS-weighted subwindows. The resulting mean reduced spacing ratios are shown in Fig.~\ref{fig:r_evolution}, where the Poisson, GOE, and GUE reference values are included as guides for comparison.

\subsubsection{Energy-Resolved Spectral Statistics}

The energy-resolved analysis reveals a considerably more structured evolution than that observed for the complete spectral window. Along the $U$ cut, the strongest level repulsion first develops in the central region of the $4f$ manifold. At $U=2.72$ eV, the central-window ratio reaches $\langle r\rangle=0.5906$. It increases further to $0.6388$ at $U=4.08$ eV and remains at $0.6349$ for $U=4.76$ eV. The latter two values are slightly larger than the GUE reference value of $0.6027$, indicating a strongly repulsive spectral regime. The Brody parameter for the central windows at $U=4.08$ and $4.76$ eV reaches its upper fitted value, $\beta_{\mathrm B}=1$, confirming the strong suppression of small nearest-neighbor spacings.

The strongest repulsion does not remain in the central region as $U$ is increased. At $U=8.16$ and $8.84$ eV, the central-window ratios decrease to $0.3858$ and $0.3880$, respectively, close to the Poisson reference value. At the same time, the lower portion of the manifold becomes strongly repulsive. Its mean reduced ratio increases to $0.5767$ at $U=8.16$ eV and reaches $0.5908$ at $U=9.52$ eV. The corresponding Brody parameters are $1.00$ and approximately $0.95$, respectively.

The upper region follows a different evolution. Its spacing ratios remain mostly below those of the central and lower regions and become particularly small at selected interaction strengths. For example, $\langle r\rangle=0.3542$ at $U=2.04$ eV. At large $U$, the upper region remains only weakly repulsive, with $\langle r\rangle=0.4200$ at $U=8.84$ eV and $0.4440$ at $U=9.52$ eV. The corresponding Brody parameters are also close to the Poisson limit.

The evolution with $U$ therefore cannot be described as a uniform increase of spectral repulsion with electronic correlation. Instead, the region with the strongest repulsion changes its position within the $4f$ manifold. It is concentrated primarily in the central region for intermediate values of $U$ and moves toward the lower part of the manifold at larger interaction strengths. The local Coulomb interaction therefore produces a redistribution of spectral correlations across energy.

The effect of $J$ is different. The complete-window values in Table~\ref{tab:dft_observables} suggest an overall approach toward Poisson-like statistics as $J$ increases, particularly around $J=0.68$--$0.82$ eV. The energy-resolved results show that this apparent behavior does not result from a uniform reduction of level repulsion.

The central portion of the manifold becomes increasingly repulsive as $J$ approaches $0.68$ eV. Its mean ratio is $0.5610$ at $J=0.27$ eV, $0.5662$ at $J=0.41$ eV, and $0.5651$ at $J=0.54$ eV. At $J=0.68$ eV it reaches $\langle r\rangle_{\mathrm{central}}=0.5828$, lying between the GOE and GUE reference means and closer to the GUE value. The Brody parameter also reaches its upper fitted value, indicating a strong suppression of small spacings in this region.

The upper portion of the same $4f$ manifold undergoes the opposite evolution. Its mean ratio decreases from $0.5464$ at $J=0.41$ eV to $0.4235$ at $J=0.54$ eV and then falls sharply to $\langle r\rangle_{\mathrm{upper}}=0.2674$ at $J=0.68$ eV. At $J=0.82$ eV, the ratio decreases slightly further to $0.2628$. These values are substantially below the Poisson mean of $0.3863$. The corresponding Brody parameters at $J=0.68$ and $0.82$ eV are approximately $0.09$ and $0.07$, respectively.

Values of $\langle r\rangle$ below the Poisson reference indicate a large contribution from sequences in which one spacing is much smaller than the adjacent spacing. The upper region around $J=0.68$--$0.82$ eV is therefore better described as a clustered spectral regime rather than simply as an uncorrelated Poisson spectrum.

The strongest contrast occurs at $J=0.68$ eV, where the central and upper regions of the same $4f$ manifold simultaneously exhibit $\langle r\rangle_{\mathrm{central}}=0.5828$ and $\langle r\rangle_{\mathrm{upper}}=0.2674$. When the complete manifold is considered, these two very different regions are combined and the resulting ratio is only $\langle r\rangle_{\mathrm{full}}=0.4122$. The apparently Poisson-like full-window statistic therefore does not represent a uniformly uncorrelated spectrum. Instead, it results from averaging spectral regions with strongly contrasting local correlations. This coexistence therefore indicates that the exchange interaction does not produce a uniform increase or decrease in spectral correlation, but instead reorganizes different regions of the same electronic manifold in opposite statistical directions.

This result is particularly significant when compared with the electronic-structure quantities in Table~\ref{tab:dft_observables}. At $J=0.68$ eV, the spin magnetic moment and total $4f$ occupation remain very close to their values at $J=0.0$. The energy-resolved spectral statistics therefore reveal an internal reorganization of the $4f$ manifold that is largely hidden in the conventional integrated observables.

\begin{figure}[!htbp]
\centering
\includegraphics[width=\textwidth]{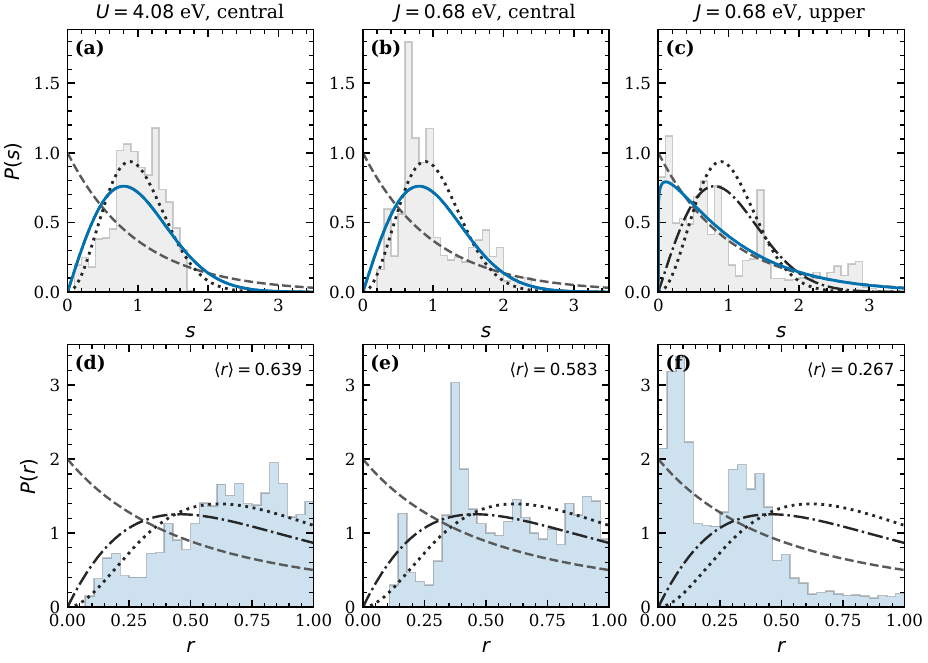}
\caption{Representative nearest-neighbor spacing distributions $P(s)$ (upper panels) and reduced spacing-ratio distributions $P(r)$ (lower panels) for selected spectral regions of the Ce-$4f$ manifold. The columns correspond to the central window at $U = 4.08$ eV ($J = 0.0$ eV), and the central and upper windows at $U = 6.12$ eV and $J = 0.68$ eV, respectively. Analytical RMT reference distributions (Poisson, GOE, and GUE) are shown alongside the corresponding Brody fits and mean reduced spacing ratios $\langle r \rangle$.}

\label{fig:representative_rmt}
\end{figure}

Representative nearest-neighbor spacing and spacing-ratio distributions are shown in Fig.~\ref{fig:representative_rmt}. The central window at $U=4.08$ eV represents the strongly repulsive intermediate-$U$ regime. The central and upper windows at $J=0.68$ eV illustrate the coexistence of strongly repulsive and clustered statistics at the same interaction parameters.

\subsubsection{Band Velocity Statistics}

The level-spacing statistics describe the relative organization of neighboring eigenvalues, but they do not directly show how the corresponding bands evolve along the reciprocal-space path. The scaled band-velocity distributions provide complementary information about the 
dispersion of the electronic states within each spectral regime.

For the complete $J$ window, the width of the scaled velocity distribution decreases gradually from approximately $12.25$ at $J=0.0$ eV to $11.18$ at $J=0.95$ eV. The distributions remain approximately centered near zero, but they contain heavier tails than a Gaussian distribution. The excess kurtosis increases from approximately $2.06$ to $3.28$ across the $J$ series, indicating that large velocity events remain more frequent than expected for an ideal Gaussian distribution.

The energy-resolved results show a different behavior. In the central region, the velocity distributions become comparatively symmetric around $J=0.68$--$0.95$ eV. At $J=0.68$ eV, the scaled velocity distribution has a fitted width of approximately $5.88$ and an excess kurtosis of $0.18$. This is the same region in which the spacing-ratio statistics show strong level repulsion.

The upper region at the same interaction strength exhibits a broader and more structured velocity distribution. Although its center remains close to zero, the distribution contains different populations of band slopes and departs more strongly from a simple Gaussian shape. The clustered spectral regime is therefore not associated only with uniformly flat bands, but with a more heterogeneous distribution of dispersions.

A similar connection appears along the $U$ cut. The central windows near $U=4.08$--$4.76$ eV, where the spacing statistics show the strongest repulsion, have comparatively symmetric velocity distributions. At larger $U$, the upper region develops a pronounced asymmetry. Its skewness is approximately $-2.32$ at $U=8.16$ eV and reaches $-3.14$ at $U=9.52$ eV, while the excess kurtosis increases from approximately $8.26$ to $14.39$. The upper high-$U$ region therefore contains a majority of states with moderate slopes together with a smaller population of bands with unusually large directional velocities.

\subsubsection{Band Curvature Statistics}

The curvature statistics provide a more direct measure of how rapidly the band slopes change along the $W\rightarrow L$ path. Because positive and negative curvatures can cancel, the analysis is based primarily on the magnitude of the scaled curvature, $|\widetilde C|$. The median represents the typical curvature scale, while the high-curvature fraction characterizes the contribution of unusually sharp events.

Along the $U$ cut, the large reduced-ratio values observed in the central region at intermediate $U$ are accompanied by large typical curvatures. At $U=4.08$ eV, the median scaled curvature in the central region is approximately $301$. Only about $1.2\%$ of the curvature values exceed five times the local median, indicating that the large curvature is distributed through a substantial part of the selected spectrum rather than being produced only by isolated sharp events.

As the strongly repulsive region moves toward the lower portion of the manifold at larger $U$, the curvature follows the same tendency. The median scaled curvature in the lower region is approximately $256$ at $U=8.16$ eV and $269$ at $U=9.52$ eV. The corresponding physical median curvature also increases, reaching approximately $1.19~\mathrm{eV\,\AA^2}$ at the largest $U$. This shows that the change is not produced only by the scaling with the local level separation, but is also present in the physical bending of the bands.

The clearest contrast between level statistics and band dynamics is again found at $J=0.68$ eV. In the central region, where $\langle r\rangle=0.5828$, the median scaled curvature is approximately $258.8$. The high-curvature fraction is about $5.5\%$, while small-gap minima occur at a rate of only approximately $0.284$ per 1000 adjacent-level and $k$-point samples.

The upper region at the same values of $U$ and $J$ presents the opposite behavior. Its median scaled curvature decreases to approximately $112.0$, while the fraction of high-curvature events increases to approximately $13.3\%$. The small-gap rate also increases strongly, reaching approximately $3.36$ events per 1000 adjacent-level and $k$-point samples.

\begin{figure}[!htbp]
\centering
\includegraphics[width=\linewidth]{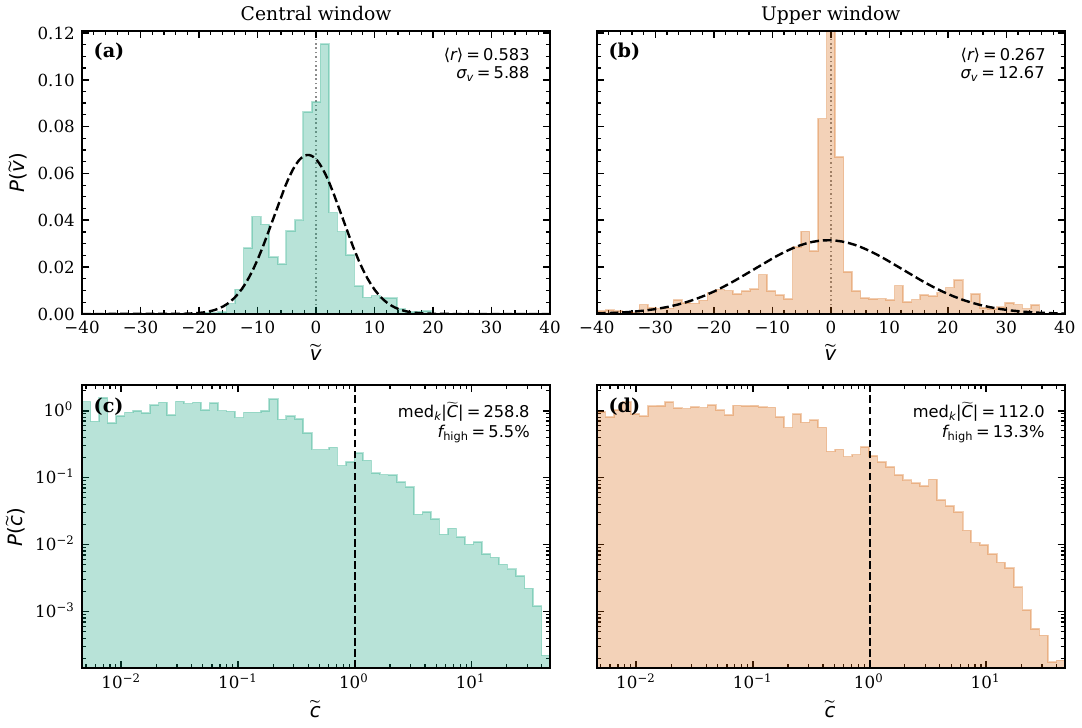}
\caption{Comparison of the band dynamics in the central (left panels) and upper (right panels) regions of the Ce-$4f$ manifold at $U = 6.12$ eV and $J = 0.68$ eV. The signed scaled-velocity distributions $P(\tilde{v})$ (upper panels) characterize the local band slopes, while the normalized curvature-magnitude distributions $P(\tilde{c})$ (lower panels) display the typical curvature scale and the high-curvature tails.}
\label{fig:band_dynamics_J005}
\end{figure}

These results indicate two qualitatively different forms of spectral organization. The strongly repulsive central region has a large typical curvature and very few near-degeneracies. In contrast, the clustered upper region contains a larger number of small gaps and has a lower typical curvature, but a greater relative contribution from isolated sharp-curvature events. This behavior is consistent with a spectrum containing many relatively weakly interacting or closely spaced bands, interrupted by a smaller number of strong local rearrangements. The corresponding velocity and curvature distributions are compared directly in Fig.~\ref{fig:band_dynamics_J005}, highlighting the contrast between the strongly repulsive central region and the clustered upper region.

\subsection{Relativistic-Character-Resolved Spectral Statistics}

The energy-resolved analysis shows that different regions of the same Ce-$4f$ manifold may exhibit strongly contrasting spectral statistics. An additional question is whether this behavior results primarily from the superposition of states with different relativistic character or whether the statistical reorganization persists within the individual components of the spin-orbit-split $4f$ manifold. The $j=5/2$ and $j=7/2$ projected components overlap substantially in energy, as shown by the pDOS in Figs.~\ref{fig:electronic_structure_U} and \ref{fig:electronic_structure_J}. The character-resolved analysis therefore does not represent an additional subdivision of the $4f$ manifold into separate energy intervals. Instead, it separates overlapping states according to their dominant relativistic character.

\begin{figure}[!htbp]
\centering
\includegraphics[width=\linewidth]{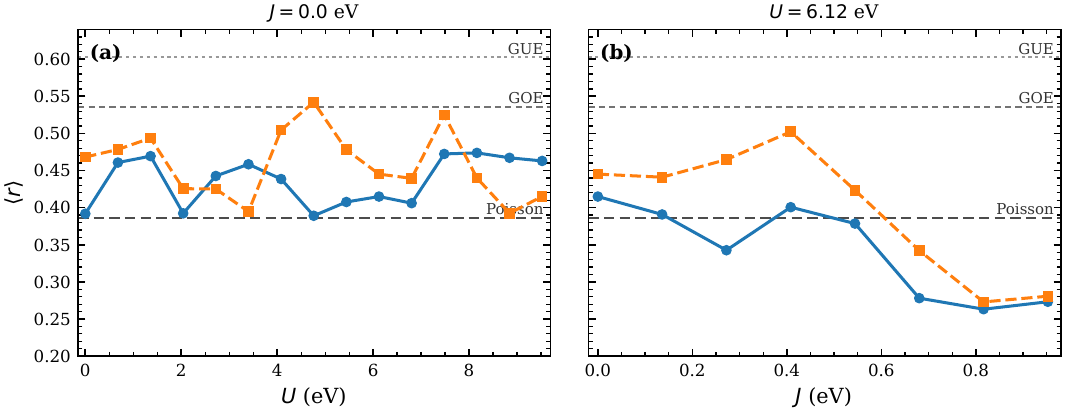}
\caption{Evolution of the mean reduced spacing ratio $\langle r \rangle$ for states with dominant $j = 5/2$ and $j = 7/2$ Ce-$4f$ character as a function of (a) Hubbard $U$ (at $J = 0.0$ eV) and (b) exchange parameter $J$ (at fixed $U = 6.12$ eV). Horizontal dashed lines represent the analytical Poisson, GOE, and GUE random matrix reference values.}
\label{fig:j_resolved_statistics}
\end{figure}

To examine this possibility, the high-energy $4f$ spectrum was resolved according to its dominant $j=5/2$ and $j=7/2$ character using the selection procedure described in Sec.~2.2. Figure~\ref{fig:j_resolved_statistics} shows the evolution of the mean reduced spacing ratio for the two relativistic subsequences along both interaction paths.

Along the $U$ series, neither relativistic component follows a monotonic evolution toward one of the RMT reference limits. The $j=5/2$ and $j=7/2$ subsequences instead display distinct nonmonotonic responses as the interaction changes. The statistical reorganization of the $4f$ spectrum therefore occurs not only across different energy regions, but also within its relativistic character.

Along the $J$ series, both character-resolved subsequences develop substantially smaller values of $\langle r\rangle$ at the larger exchange interactions. At $J=0.68$ eV, the mean ratios are $\langle r\rangle_{5/2}=0.2781$ and $\langle r\rangle_{7/2}=0.3422$. Both values lie below the Poisson reference mean, while the corresponding Brody parameters decrease to $\beta_{\mathrm B}=0.188$ and $0.126$, respectively. At $J=0.82$ eV, the ratios decrease further to $0.2632$ and $0.2730$. The clustering observed in the upper part of the complete $4f$ manifold therefore persists after the spectrum is separated according to its dominant relativistic character. It cannot be attributed only to the superposition of $j=5/2$ and $j=7/2$ levels.

The character selection reduces the number of usable reciprocal-space points more strongly for the $j=5/2$ subsequence. For the representative cases $U=5.44$ eV, $J=0.0$; $U=6.12$ eV, $J=0.0$; and $U=6.12$ eV, $J=0.68$ eV, the $j=5/2$ analysis retains 859, 779, and 634 of the 901 reciprocal-space points, respectively. Along the $J$ series this number decreases further to 402 points at $J=0.82$ eV and 294 points at $J=0.95$ eV. In contrast, the $j=7/2$ subsequence retains all 901 points throughout the $J$ cut. The high-$J$ $j=5/2$ values should therefore be interpreted with the reduced sampling in mind, while the corresponding $j=7/2$ tendency is obtained over the complete reciprocal-space interval.

The character-resolved statistics for three representative interaction pairs are summarized in Table~\ref{tab:j_resolved}. The first and third interaction pairs have the same effective interaction, $U-J=5.44$ eV, while the second and third have the same Coulomb interaction, $U=6.12$ eV.

\begin{table}[H]
\centering
\caption{Character-resolved spectral statistics for representative interaction parameters. States were assigned using $w_f\geq0.05$ and a minimum $j$-character purity of $0.60$. $\langle r\rangle$ is the mean reduced consecutive-spacing ratio and $\beta_{\mathrm B}$ is the Brody parameter obtained from an unbinned maximum-likelihood fit. The first and third interaction pairs have the same $U-J=5.44$ eV, while the second and third have the same $U=6.12$ eV.}
\label{tab:j_resolved}
\small
\setlength{\tabcolsep}{4pt}
\begin{tabular}{cccccc}
\hline
$U$ (eV) & $J$ (eV) & $U-J$ (eV) & $j$ & $\langle r\rangle$ & $\beta_{\mathrm B}$ \\
\hline
5.44 & 0.00 & 5.44 & $5/2$ & 0.408 & 0.550 \\
5.44 & 0.00 & 5.44 & $7/2$ & 0.478 & 0.663 \\
6.12 & 0.00 & 6.12 & $5/2$ & 0.415 & 0.725 \\
6.12 & 0.00 & 6.12 & $7/2$ & 0.445 & 0.547 \\
6.12 & 0.68 & 5.44 & $5/2$ & 0.278 & 0.188 \\
6.12 & 0.68 & 5.44 & $7/2$ & 0.342 & 0.126 \\
\hline
\end{tabular}
\end{table}

\begin{figure}[!htbp]
\centering
\includegraphics[width=\linewidth]{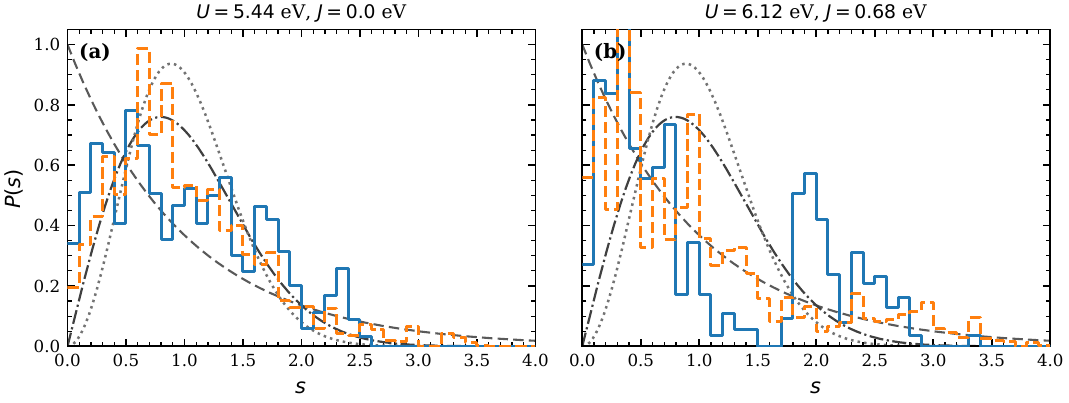}
\caption{Character-resolved nearest-neighbor spacing distributions $P(s)$ for two calculations that share the same effective value $U - J = 5.44$ eV: (a) $U = 5.44$ eV, $J = 0.0$ eV and (b) $U = 6.12$ eV, $J = 0.68$ eV. Blue solid and orange dashed lines represent the $j = 5/2$ and $j = 7/2$ subsequences, respectively, shown alongside the analytical RMT reference distributions (grayscale).}
\label{fig:equal_ueff_j}
\end{figure}

A particularly useful comparison is obtained from the two calculations with the same numerical value $U-J=5.44$ eV. For $U=5.44$ eV and $J=0.0$, the character-resolved ratios are $\langle r\rangle_{5/2}=0.4076$ and $\langle r\rangle_{7/2}=0.4780$. When the same value of $U-J$ is obtained using $U=6.12$ eV and $J=0.68$ eV, these values decrease to $0.2781$ and $0.3422$, respectively. The corresponding nearest-neighbor spacing distributions are compared in Fig.~\ref{fig:equal_ueff_j}. Both relativistic subsequences acquire substantially greater weight at small spacings when the exchange interaction is included explicitly.

The comparison at fixed $U$ leads to the same conclusion. At $U=6.12$ eV, increasing $J$ from $0.0$ to $0.68$ eV changes $\langle r\rangle_{5/2}$ from $0.4150$ to $0.2781$ and $\langle r\rangle_{7/2}$ from $0.4452$ to $0.3422$. The statistical response of the $4f$ spectrum therefore cannot be described solely through the scalar combination $U-J$. Two calculations with the same numerical effective interaction retain substantially different short-range spectral correlations when the exchange interaction is treated explicitly. Within the rotationally invariant DFT+$U$+SOC framework used here, the explicit exchange interaction therefore contributes to the internal organization of the correlated $4f$ spectrum in a way that is not captured by the scalar combination $U-J$ alone.

The character-resolved analysis therefore extends the energy-resolved result. A complete-manifold statistic may average over different energy regions, but it may also combine states with different relativistic character. Resolving both contributions shows that the exchange-driven clustering is an internal property of the calculated $4f$ spectrum rather than only a consequence of combining distinct spectral subsequences.

\subsection{Interpretation and Scope of the Statistical Analysis}

Taken together, the results indicate two distinct effects of the local interaction: a Coulomb-driven migration of spectral repulsion across energy and an exchange-driven coexistence of repulsive and clustered spectral sectors. The energy- and character-resolved results demonstrate that a statistical characterization of the complete Ce-$4f$ manifold can conceal important differences in its internal organization. This is most evident at $U=6.12$ eV and $J=0.68$ eV, where the central and upper energy windows exhibit strongly contrasting reduced spacing ratios, while their combined full-window value lies much closer to the Poisson reference. The relativistic-character analysis further shows that the clustering at this interaction strength persists within both the $j=5/2$ and $j=7/2$ subsequences. The complete-window statistic therefore represents an average over spectral structure that is heterogeneous both across energy and with respect to the dominant relativistic character of the states.

This internal reorganization is only weakly reflected in the conventional electronic-structure observables. Along the $J$ series, the total spin magnetic moment and Ce-$4f$ occupation remain nearly constant even where the energy-resolved spacing statistics change substantially. Treating the calculated spectrum as an energy-resolved statistical object therefore provides information that is not evident from the integrated DFT quantities alone.

The velocity and curvature distributions show that the contrasting spectral regimes are also accompanied by different forms of band dynamics. The strongly repulsive central region has a narrower scaled-velocity distribution, a larger typical curvature, and a suppression of small-gap events. The clustered upper region instead combines a broader and more structured velocity distribution with a smaller typical curvature, more frequent small gaps, and a larger relative contribution from sharp curvature events. The spacing, velocity, and curvature statistics thus describe complementary aspects of the same spectral reorganization.

The robustness of these conclusions with respect to the reciprocal-space trajectory was examined by repeating the short-range statistical analysis along $L\rightarrow K$. The detailed values of $\langle r\rangle$ show a directional dependence, as expected for spectra sampled along different regions of reciprocal space, but the principal redistribution of spectral correlations is preserved. At large $U$, the relative enhancement of spectral correlations shifts toward the lower portion of the $4f$ manifold on both trajectories. At $J=0.68$~eV, the strongly clustered upper region is also reproduced, with $\langle r\rangle_{\mathrm{upper}}=0.2756$ along $L\rightarrow K$, compared with $0.2674$ along $W\rightarrow L$. The central--upper contrast remains evident along $L\rightarrow K$, although the central-window ratio decreases from $0.5828$ to $0.4633$, demonstrating that the absolute strength of the level repulsion is direction dependent. The exchange-driven clustering within the relativistic subsequences is likewise preserved: the $L\rightarrow K$ calculation gives $\langle r\rangle_{5/2}=0.2291$ and $\langle r\rangle_{7/2}=0.2917$, compared with $0.2781$ and $0.3422$,  respectively, along $W\rightarrow L$. Thus, the redistribution of spectral correlations across energy and relativistic character is robust, while  the absolute magnitude of the short-range statistics within individual vectors retains a dependence on reciprocal-space direction.

The influence of competing self-consistent DFT+$U$ solutions was examined explicitly through the alternative-initialization analysis. Isolated points along the $U$ series were found to be sensitive to the initial electronic configuration. The clearest case occurs at $U=1.36$~eV, where an alternative converged solution lies approximately $0.626$~eV below the initially obtained state after the inclusion of spin-orbit coupling. The lower-energy solution, retained in Table~1 and Figs.~3 and~6, gives $\langle r\rangle_{\mathrm{central}}=0.5535$, compared with $0.3662$ for the higher-energy solution. Despite this sensitivity at isolated interaction values, the broader energy-resolved evolution along the $U$ series is preserved. The $J$ series is considerably less sensitive to the initialization procedure: at $J=0.68$~eV, for example, the production central and upper values $\langle r\rangle=0.5828$ and $0.2674$ change only to $0.5952$ and $0.2668$, respectively, under the alternative initialization. The exchange-driven clustering within the relativistic subsequences is similarly preserved. The principal spectral trends therefore do not depend on a particular initialization procedure, although individual points along the $U$ series may access competing self-consistent states.

The RMT ensembles are used here as statistical reference distributions rather than as exact symmetry assignments of the calculated Hamiltonian. The analyzed eigenvalues form an effective one-electron spectrum obtained within a static spin-polarized DFT+$U$+SOC framework and do not represent the complete many-body excitation spectrum or a full description of the experimental paramagnetic $\gamma$ phase of Ce. In addition, the derivatives were calculated using energy-ordered eigenvalue indices rather than wavefunction-overlap tracking. The largest curvature events may therefore include contributions from avoided crossings, exact crossings, or changes in the energy ordering of neighboring bands. The median curvature is consequently used as the principal measure of the typical band response, while the extreme tail is interpreted more cautiously as an indicator of crossing and band-reordering activity.

%Part of the nonmonotonic dependence of the spectral statistics on $U$ and $J$ may also be related to the multiminimum character of the DFT+$U$ energy functional for the Ce-$4f$ shell. Correlated-band calculations for $\gamma$-Ce have previously identified multiple self-consistent ground and metastable electronic configurations \cite{shick2001cerium}. Because each $(U,J)$ point in the present work was converged independently, successive interaction values were not constrained to remain on the same self-consistent solution branch. Consequently, some abrupt or nonmonotonic changes may reflect crossings or switching between nearby local minima with different $4f$ occupation matrices, in addition to a continuous reorganization of the bands within a single branch. The present calculations do not systematically separate these two contributions, and the nonmonotonic behavior is therefore interpreted as a property of the independently converged effective one-electron spectra.

%This possibility is consistent with the observation that substantial changes in the energy- and character-resolved spectral statistics can occur while the total $4f$ occupation and spin magnetic moment remain comparatively stable. Nearby self-consistent solutions may differ in the internal orbital organization and dispersion of the $4f$ bands without producing equally large changes in these integrated quantities.

% ------------------------------------- CONCLUSIONS --------------------------------------------

\section{Conclusions}\label{Sec:Conclusions}

The electronic spectrum of $\gamma$-Ce was examined using spin-polarized DFT+$U$+SOC calculations combined with energy-resolved spectral statistics. The DOS-dependent selection of the high-energy Ce-$4f$ manifold allowed the same physically relevant spectral region to be followed as its position and width changed with the Coulomb and exchange interactions.

The results show that electronic correlation does not produce a single monotonic transition between Poisson-like and Wigner--Dyson-like statistics. Increasing $U$ changes the location of the most strongly repulsive region within the $4f$ manifold. At intermediate $U$, the strongest repulsion occurs in the central region, whereas at larger $U$ it develops in the lower part of the manifold. The Coulomb interaction therefore redistributes spectral correlations across energy.

The exchange interaction produces a different reorganization. At fixed $U=6.12$ eV and near $J=0.68$ eV, the central and upper regions simultaneously exhibit strongly repulsive and clustered statistics, respectively. Their reduced spacing ratios, $\langle r\rangle_{\mathrm{central}}=0.5828$ and $\langle r\rangle_{\mathrm{upper}}=0.2674$, combine to produce a full-window value of $\langle r\rangle_{\mathrm{full}}=0.4122$. A global spectral statistic can therefore conceal the coexistence of distinct regimes within the same electronic manifold.

The relativistic-character-resolved analysis shows that the exchange-driven clustering is not produced only by the superposition of the $j=5/2$ and $j=7/2$ components. At $U=6.12$ eV and $J=0.68$ eV, both subsequences develop strongly clustered statistics. Moreover, comparison with the $U=5.44$ eV, $J=0$ calculation, which has the same numerical value $U-J=5.44$ eV, produces substantially different character-resolved spacing statistics. The separate treatment of Coulomb and exchange interactions therefore modifies the internal organization of the relativistic $4f$ subsequences in a manner that cannot be represented completely by a single scalar effective interaction $U-J$.

The band-velocity and curvature analysis shows that the contrasting spectral regimes are accompanied by different forms of band dynamics. At $J=0.68$ eV, the strongly repulsive central region has a comparatively narrow scaled-velocity distribution and a larger typical curvature, whereas the clustered upper region combines a broader velocity distribution, a smaller typical curvature, more frequent small gaps, and a larger high-curvature tail. These statistical changes occur while the total spin magnetic moment and Ce-$4f$ occupation remain nearly unchanged along much of the $J$ series. Energy-resolved and character-resolved spectral statistics consequently reveal correlation-driven reorganization that is not captured by conventional integrated DFT observables or by averages over the complete spectral window.

The principal conclusions were also tested against the choice of self-consistent initialization and the reciprocal-space trajectory. Alternative initializations revealed competing DFT+$U$ solutions at isolated interaction strengths, for which the lower-energy solution was retained for the reported results, while the broader redistribution of spectral correlations was preserved. Repeating the short-range analysis along $L\rightarrow K$ reproduced the large-$U$ shift of the relative spectral enhancement toward the lower part of the $4f$ manifold, the exchange-driven clustering of the upper spectral region, and the clustering within both relativistic subsequences. Although the absolute values of the short-range statistics remain dependent on reciprocal-space direction, the principal energy- and character-resolved trends are therefore not specific to the $W\rightarrow L$ trajectory.

Within the scope of the effective one-electron DFT+$U$+SOC description used here, the evolution of $\gamma$-Ce is therefore governed not by a uniform statistical crossover but by a redistribution of spectral correlations across energy and relativistic character. The Coulomb and exchange interactions play distinguishable roles in this process, and their effects may remain hidden in conventional integrated electronic-structure observables. More broadly, these results suggest that energy- and character-resolved spectral statistics can provide a sensitive diagnostic of internal electronic reorganization in correlated band structures, particularly when different interaction terms produce similar conventional observables but different microscopic spectral organization.

% ------------------------------------- IOP END MATTER ----------------------------------------------------------------

\section*{Acknowledgments}
We acknowledge UDLAP for providing the facilities for this investigation. The authors report no conflict of interest.

\section*{Funding}
This research received no specific grant from any funding agency in the public, commercial, or not-for-profit sectors.

\section*{Data availability}
The data supporting the findings of this study are not publicly available because they form part of ongoing related research. The data required to support the results reported in this article are available from the corresponding author upon reasonable request.
% IOP requires a reason when supporting data are not publicly available. Add the real reason in the submission system and, if requested, in this statement before final submission.

% ------------------------------------- BIBLIOGRAFIA ----------------------------------------------------------------

%\section{References}\label{Referencies}
% \nocite{*}
\bibliographystyle{unsrt}
\bibliography{chaos_bibliography}

\end{document}